\documentclass{emulateapj}

\usepackage{graphicx}
\usepackage{amssymb}
\usepackage{amsmath}
\usepackage[usenames]{color}
\usepackage{times}

\newcommand{\msun}{M_\odot}
\newcommand{\mstar}{M_\ast}
\newcommand{\pc}{\mathrm{pc}}
\newcommand{\mbh}{M_\bullet}

\newcommand{\model}[1]{{\tt{}#1}}
\newcommand{\myr}{\mathrm{Myr}}
\newcommand{\rd}{\mathrm{d}}

\newcommand{\mas}{^{\prime\prime}}
\newcommand{\bs}{}
\newcommand{\df}{{\cal N}}
\newcommand{\iH}{i_\mathrm{H}}
\newcommand{\eH}{e_\mathrm{H}}

\newcommand{\mean}[1]{\langle #1 \rangle}

\shorttitle{Relaxation of the young stellar disc in the GC}
\shortauthors{\v{S}ubr \& Haas}

\begin{document}
\title{Two-body relaxation
driven evolution of the young stellar disc in the Galactic Centre}
\author{Ladislav \v{S}ubr$^{1}$ and Jaroslav Haas$^{1}$}
\affil{$^{1}$Charles University in Prague, Faculty of Mathematics and Physics,
Astronomical Institute, V Hole\v{s}ovi\v{c}k\'ach 2, Praha, CZ-18000,
Czech Republic}

\begin{abstract}
The centre of our Galaxy hosts almost two hundreds of very young stars,
a subset of which is orbiting the central supermassive black hole (SMBH) in
a relatively thin disc-like structure. First analyses indicated a power-law
surface density profile of the disc, $\Sigma \propto R^\beta$ with
$\beta = -2$. Recently, however, speculations about this
profile arose. In particular, it now seems to be better described by a
sort of broken power-law. By means of both analytical arguments and
numerical $N$-body modelling, we show that such a broken power-law profile is a
natural consequence of the two-body relaxation of the disc which is, due
to small relative velocities of stars on nearby co-planar Keplerian
orbits around the SMBH, effective enough to affect the evolution of the
disc on time-scales comparable to its estimated age. In the inner,
densest part of the disc, the profile becomes rather flat ($\beta\approx-1$)
while the outer parts keep imprints of the initial state. Our numerical models
show that the observed projected surface density profile of the young stellar
disc can result from two-body relaxation driven evolution of a disc with
initial single power-law profile with $-2\lesssim\beta\lesssim-1.5$.
In addition, we suggest that two-body relaxation may have caused a
significant radial migration of the S-stars towards the central SMBH,
thus playing an important role in their formation scenario.
\end{abstract}
\keywords{Galaxy: nucleus -- methods: numerical -- stars: kinematics and dynamics}
\section{Introduction}
It has already become a text book level piece of knowledge that galactic nuclei
in general contain an extended cluster of old stars
surrounding the central supermassive black hole (SMBH). The properties of such
a star cluster in the Galactic Centre (GC) have been discussed many times
in the literature \citep[e.g.][]{bse2009,do2009,do2013} as well as the mass
of the SMBH \citep[$\approx4\times10^6\msun$; e.g.][]{ghez2008} and its
distance from the Sun \citep[$\approx 8\,\mathrm{kpc}$; e.g.][]{gillessen2009}.
Improvement of observational instruments and techniques in the past few
decades, however, led to a surprising discovery of a significant number of
young stars in the innermost parsec of the Galaxy
\citep[e.g.][]{paumard2006,bartko2009,bartko2010,do2013} which appear to be
orbiting the central SMBH on near-Keplerian orbits. Those in the immediate
vicinity of the SMBH, the so-called S-stars, have been identified as quite usual,
$\approx20\,\myr$ old stars of spectral type B with randomly oriented
orbits \citep[e.g.][]{ghez2005}. Stars that are observed farther out (projected
distance $\gtrsim 1\mas$) have been identified as massive, $\approx6\,\myr$ old
OB- or WR-stars \citep[e.g.][]{paumard2006}. Contrary to the S-stars, it has been
suggested \citep{lb2003} that a non-negligible subset of these massive stars
form a coherently rotating stellar structure around the SMBH -- a stellar disc,
which is commonly called the clockwise system (CWS). This exciting possibility
inspired many authors to study various aspects of the dynamical evolution of
such stellar discs.

Attention has been mostly paid to the evolution of eccentricities and
inclinations of the stellar orbits in the disc \citep[e.g.][]{aba2007,caa2008}
due to the still ongoing debate about the origin of those young
stars that are observed off the suggested disc plane. Evolution of another
important quantity, semi-major axes of the orbits in the disc, which determines
its radial structure has been, however, usually neglected. Among others,
overlooking this question might have had the two following causes. First, raw
estimates of the classical two-body relaxation time, which is supposed to be
adequate for changes of orbital energy, gave, considering the parameters of the
old star cluster, orders of magnitude larger values than the age of
the young stars. In contrast, the characteristic time-scale of resonant
relaxation (either vector or scalar) that are relevant for changes of angular
momentum and, consequently, eccentricity and inclination is estimated to be
of the order of millions of years. Second, analyses of
the observational data usually led to a statement that the surface (number)
density profile of the CWS is compatible with a single power-law $\Sigma
\propto R^\beta$ with $\beta = -2$ \citep{paumard2006,lu2009} which is widely
accepted to be a natural outcome of the likely formation scenario of such
stellar discs \citep{lp1987,levin2007} and, therefore, there seems nothing
interesting to be done.

Nevertheless, most recently, pieces of evidence that a broken power-law surface
density profile may better fit the observational data have been reported.
\cite{bse2009} find values $\beta=-1.08$ and $-3.46$ with a break-point
at $0.4\,\pc\; (10\mas)$ for which they do not provide any explanation.
Furthermore, although not giving explicit values of the broken power-law
indices, \cite{do2013} find a sort of plateau in their surface density profile
which is followed by a sharp drop of density beyond $0.16\,\pc\; (4\mas)$. Is
there any dynamical reason for a broken power-law to be a better description
of an evolved stellar disc around a SMBH? In this paper, we address this question.
\section{Theoretical considerations}
\label{sec:theory}
A classical formula for two-body relaxation time within a self-gravitating
system of stars of mass $\mstar$, number density $n$ and velocity dispersion
$\sigma$ reads \citep{bt1987}:
\begin{equation}
 t_\mathrm{R} \approx 0.34\,\frac{\sigma^3}{G^2 \mstar^2\, n \ln\Lambda}\;,
\label{eq:t_relax}
\end{equation}
where $G$ stands for the gravitational constant and $\ln \Lambda$ is the
Coulomb logarithm. Formula~(\ref{eq:t_relax})
gives values of the order of $10^9\,\mathrm{yr}$ for the old star cluster
in the GC when replacing $\sigma$ with Keplerian orbital velocity around
the SMBH, $\sigma \approx v_\mathrm{K} \equiv \sqrt{G\,\mbh / R}$:
\begin{eqnarray}
 t_\mathrm{R}(R) &\bs\approx  \bs& \frac{0.2}{\ln\Lambda}
 \left( \frac{N(R)}{10^5} \right)^{-1}
 \left( \frac{\mbh}{4\times 10^6\msun} \right)^{-1/2}\bs\times \nonumber\\
 & &\left(\frac{\mbh / \mstar}{10^6} \right)^2
 \left( \frac{R}{0.04\,\pc} \right)^{3/2} \mathrm{Gyr}\;.
\end{eqnarray}
Here, $N(R) \approx n\,R^3$ denotes the number of stars below radius $R$.
The situation is, however, different if we consider a thin stellar disc
whose density (both mass and number) exceeds the density of the old nuclear
star cluster. At the same time, and more importantly,
relative velocities of stars within the coherently rotating disc are much
smaller than their orbital velocity. Both these facts lead to a considerable
(orders of magnitude) shortening of the two-body relaxation time-scale within
a thin stellar disc with respect to what could be inferred from the
parameters of the old star cluster.

The theory of relaxation of thin discs of bodies driven by their mutual
gravitational interaction has already been well established in the context
of protoplanetary discs. Let us thus forward the reader to the paper
\cite{si2000} which summarises different theoretical approaches to the problem.
The authors derive equation (4.10) for temporal evolution of the mean square
eccentricity which can be rewritten in the form
\begin{equation}
 \frac{\rd \mean{e^2}}{\rd t} = \frac{\mean{e^2}}{T_\mathrm{char}}
 \makebox[4em]{with} T_\mathrm{char} = \frac{0.06}{\alpha}
 \frac{\mean{e^2}^2}{\Omega\, \Sigma\, R^2}
 \left( \frac{\mbh}{M_\ast} \right)^2\;,
\label{eq:dedt}
\end{equation}
where $M_\ast$ is the mass of individual bodies in the disc (stars in our case),
$\mbh$ is the mass of the central body which dominates the gravitational potential
(the SMBH), $\Sigma$ is the surface number density of the
(stellar) disc and $\Omega \equiv \sqrt{G\,\mbh / R^3}$ is the Keplerian
angular velocity;
$\alpha$ is a numerical factor of the order of unity which can be tuned by
comparison with $N$-body integration. \cite{si2000} report $\alpha = 2$
as a suitable choice and we will adopt this value hereafter.

$T_\mathrm{char}$ introduced in (\ref{eq:dedt}) approximately
corresponds to the time required for $\mean{e^2}$ to grow
by a factor of two and, due to its dependence upon $\mean{e^2}$, it may be
arbitrarily short. A better sense of the characteristic time-scale is obtained
by integration of equation (\ref{eq:dedt}) which gives the time for mean square
eccentricity to grow from initial (nearly zero value) to $\mean{e^2}$:
\begin{equation}
 t(\mean{e^2}) = \int_0^{\mean{e^2}} \frac{\rd t}{\rd \mean{\epsilon^2}}
 \,\rd\mean{\epsilon^2} = 0.015 \frac{\mean{e^2}^2}{\Omega\,\Sigma\,R^2}
 \left( \frac{\mbh}{M_\ast} \right)^2 \;.
\label{eq:te} 
\end{equation}

For a stellar disc of total
mass $M_\rd$ and radial surface density profile $\Sigma\propto R^{-2}$ ranging
from $R_\mathrm{in}$ to $R_\mathrm{out}$, formula (\ref{eq:te}) gives
\begin{eqnarray}
 t(\mean{e^2}) &\bs\approx\bs& 1.2
 \ln\!\left(\frac{R_\mathrm{out}}{R_\mathrm{in}}\right)
 \left( \frac{R}{0.04\,\pc} \right)^{\!3/2}\!
 \left( \frac{\mbh/M_\ast}{10^{6}} \right)
 \bs\!\!\times \nonumber \\
 & & \!\!\left( \frac{\mbh/M_\rd}{1000} \right)
 \left( \frac{\mbh}{4\!\times\!10^6\msun} \right)^{\!-1/2}
 \left( \frac{\mean{e^2}}{0.01} \right)^{\!2}\! \myr\;.
\label{eq:te2}
\end{eqnarray}
We see that at least for the inner parts of the young stellar disc in the
Galactic Centre, the characteristic time for change of orbital elements
is comparable to its age.

One of the methods discussed by \cite{si2000} that
led to derivation of formula (\ref{eq:dedt}) is based on evaluation of
diffuse coefficients in the Fokker-Planck equation, quadratic in velocity.
These were then transformed to changes in square of eccentricity and
inclination. As the orbital energy is
also a quadratic form of velocity, it is reasonable to expect that the
characteristic time for evolution of the radius of the orbit
is similar to that
expressed in eqs. (\ref{eq:te}) and (\ref{eq:te2}). Consequently, we expect
that radial density profile of the stellar disc in the GC may have changed
during its life-time. Let us note that our assumption about the
evolution of semi-major axes goes beyond the analytical work of \cite{si2000}
and it requires justification by a suitable numerical model which we present
below.

The rate of radial migration has to be a function of the radial density profile
which determines strength of the two-body relaxation. A stationary
solution requires a bulk accretion rate independent of radius $R$, i.e.
$\Sigma\,a\,(\rd a/\rd t) = \mathrm{const.}$ One possible candidate for the stationary
disc density profile is the \cite{bw1976} solution with distribution function
of semi-major axes $\df_a\propto a^{1/4}$ which implies
$\Sigma(R) \propto R^{-3/4}$. In spite of that this solution was derived
under the assumption of spherical symmetry, underlying energetic balance arguments
are not affected by this assumption. Bearing this caveat in mind we may
further speculate that due to rather strong radial dependence of the
characteristic time, the stationary solution would settle in the inner parts
of the disc, while the outer parts would still keep imprints of the initial
state. Hence, for initial $\Sigma(R)\propto R^{-2}$ which clearly differs
from the expected stationary state, we would obtain a broken power-law
surface density profile of the stellar disc.
\section{Numerical Model}
\label{sec:model}
In order to investigate evolution of the disc density profile, we set up
a model integrated numerically by means of direct $N$-body integration code
\model{NBODY6} \citep{aarseth03}. With respect to the original publicly
available version of the code, we added a fixed Keplerian potential representing the
central SMBH of mass $4\times 10^6\msun$. Young stellar disc is represented
by 1200 point-like particles with mass
function $\df_{M_\ast} \propto M_\ast^{-1.5}$ in the range $\langle 1\,\msun,
\;150\,\msun\rangle$; the total mass of the disc is $M_\rd = 14700\,\msun$.
The adopted values of parameters of the mass function are motivated by recent work
of \cite{lu2013} who investigated properties of the population of young stars
in the GC. The stars were initially set on circular orbits with
normalised angular momentum vectors distributed uniformly with maximum deviation
of $1\degr$ from the mean (i.e. the initial disc opening angle is $2\degr$).
We investigated models with initial power-law surface density profile
$\Sigma(R) \propto R^\beta$ with several different values of index
$\beta \in \langle -3,\; -1\rangle$. The radial extent of the disc is set
to an interval $\langle 0.04\,\pc,\, 0.4\,\pc \rangle$.

In order to clearly distinguish general trends from random fluctuations, we set
up typically 20 numerical realisations of each model, showing
results averaged over all realisations below. The reliability of our numerical
approach has been tested by comparison with qualitatively similar models
of \cite{si2000} -- see Appendix~\ref{sec:appendix}.
\section{Results}
\begin{figure}
\begin{center}
\includegraphics[width=\columnwidth]{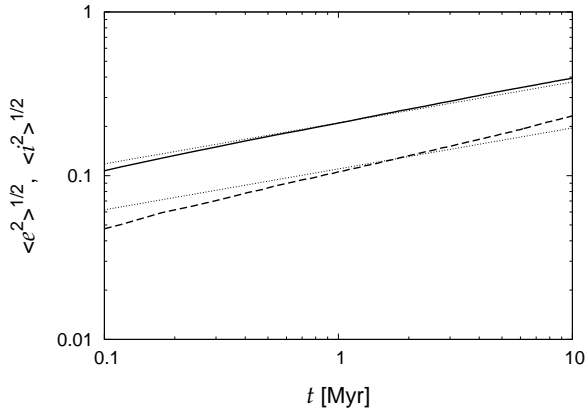}
\end{center}
\caption{Temporal evolution of the mean values of square of orbital eccentricity
(solid line) and inclination (dashed line) of the stellar orbits in the disc
for $\beta = -2$.
Thin dotted lines show dependency $t^{1/4}$ for comparison.}
\label{fig:mean_ei2}
\end{figure}
We followed dynamical evolution of the stellar disc to $t > 10\,\myr$. In terms
of the mean (square) eccentricity and inclination, the evolution is similar for
all models (which differ by the value of index $\beta$). In
Figure~\ref{fig:mean_ei2}, we show results for $\beta = -2$ which
corresponds to commonly accepted theoretically plausible surface density
profile of a stellar disc formed by self-gravitational fragmentation of
parent gaseous disc. 
Both mean values of squared eccentricity and inclination, $\mean{e^2}$ and
$\mean{i^2}$, gradually grow, i.e. the velocity dispersion within the disc evolves
towards isotropy and the disc becomes geometrically thicker. At later stages,
evolution of eccentricity follows theoretically predicted relation
$\mean{e^2}^{1/2} \propto t^{1/4}$ \citep[eq.~\ref{eq:te}, see also][]{si2000}.
According to our numerical
model, inclinations evolve somewhat faster than eccentricities and the
expected relation $\mean{i^2}^{1/2} \approx 0.5\, \mean{e^2}^{1/2}$ holds
only approximately. This apparent discrepancy is due to that the theoretical
prediction does not consider temporal evolution of the disc surface density.
\cite{si2000} have shown (and we have also verified on appropriate models --
see Appendix~\ref{sec:appendix})
that after renormalisation to constant surface density, evolution of the mean
square eccentricity and inclination matches the theoretical lines much better. The
renormalisation to constant density is, however, not possible for models with
radially dependent surface density which differs by more than an order of
magnitude between the inner and outer edge.

\begin{figure}
\begin{center}
\includegraphics[width=\columnwidth]{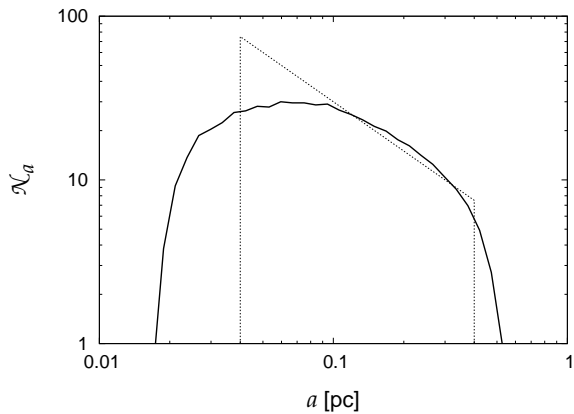}
\end{center}
\caption{Distribution of semi-major axes of the stellar disc at $t=6\,\myr$
for model with initial distribution $\df_a \propto a^{-1}$, i.e. $\Sigma(R)
\propto R^{-2}$ (solid line). Thin dotted line shows the initial state.}
\label{fig:na-2}
\end{figure}
In contrary to inclinations and eccentricities, moments of the distribution
of semi-major axes do not provide so clear view on temporal evolution of the
system. This is mainly
due to already initial order of magnitude spread of semi-major axes which
even grows in time -- few stars become unbound, i.e. reaching negative values of
semi-major axis, few others are scattered to loosely bound orbits with
$a > 1\,\pc$. We therefore plot the distribution function, $\df_a$,
of semi-major axes itself at $t = 6\,\myr$ for the model with
$\beta=-2$ in Figure~\ref{fig:na-2}. The evolved distribution
function clearly cannot be fitted by a single power-law. For $a\in \langle
0.1\,\pc,\, 0.4\,\pc \rangle$ it is still close to the initial shape,
$\df_a\propto a^{-1}$, while in the range $\langle 0.04\,\pc,\, 0.1\,\pc
\rangle$ the distribution function is roughly constant (cf. the
\citeauthor{bw1976} solution, $\df_a\propto a^{1/4}$).

The above introduced numerical model gives $\sqrt{\mean{e^2}} \approx 0.3$
at $t=6\,\myr$ (which is the estimated age of the young stars in the GC).
For the parameters of the model and this value of the mean eccentricity,
formula~(\ref{eq:te2}) gives characteristic time $t(\mean{e^2}) \approx
20\,\myr$ at the disc inner edge if we replace $\mstar$ with $\mean{\mstar}$
which equals $\approx 12\,\msun$ in our case.
The fact that the numerical model shows evident disc
evolution up to $R \approx 0.1\,\pc$ already at $t \approx 6\,\myr$
is likely due to the broad mass spectrum which is known to accelerate the
two-body relaxation driven evolution of self-gravitating systems.

\begin{figure}
\begin{center}
\includegraphics[width=\columnwidth]{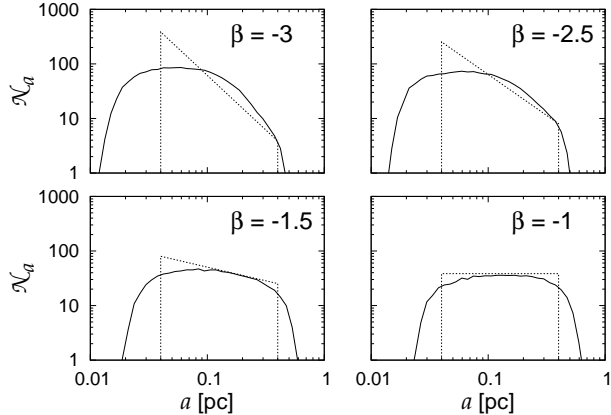}
\end{center}
\caption{Arbitrarily normalised distributions of semi-major axes for models
with different initial
surface density profiles, $\Sigma \propto R^\beta$, as indicated in each panel.
Solid lines show evolved states at $t=6\,\myr$, while thin dotted lines
represent the initial states.}
\label{fig:na}
\end{figure}

Figure~\ref{fig:na} demonstrates that the flattening of the distribution
of the semi-major axes in the inner parts of the disc holds for a wide range
of initial values of index $\beta$. This suggests a possibility that there
exists an equilibrium state, towards which the two-body relaxation of a thin
stellar disc around a dominating central mass leads. The equilibrium state
inferred from our numerical models, $\df_a \approx \mathrm{const.}$, appears
to be close to the \cite{bw1976} solution mentioned
in Section~\ref{sec:theory}. Due to the radial dependence of the relaxation
time, the flat distribution of semi-major axes settles first in the inner
parts of the disc and gradually propagates outwards.
 In the outer parts of the disc, where the two-body relaxation
time-scale is still longer than the evolution time, the distribution is
affected much less.

A generic consequence of the shear which acts in differentially rotating
discs is their radial spreading (both inward and outward). We attribute two
features of the distribution
shown in Figure~\ref{fig:na} to this process: first, it is the steepening of the
initial density profile, visible particularly for model with $\beta = -3$
beyond $R\approx0.1\,\pc$, which is due to stars that migrated outward from the
inner, very dense region. Second, as the rate of radial migration is proportional
to the disc density, we see more prominent inward migration of stars across the
initial disc inner radius for steeper density profiles (lower $\beta$). We will discuss
this effect in Section~\ref{sec:conclusions} in the context of the S-stars.

\begin{figure}
\begin{center}
\includegraphics[width=\columnwidth]{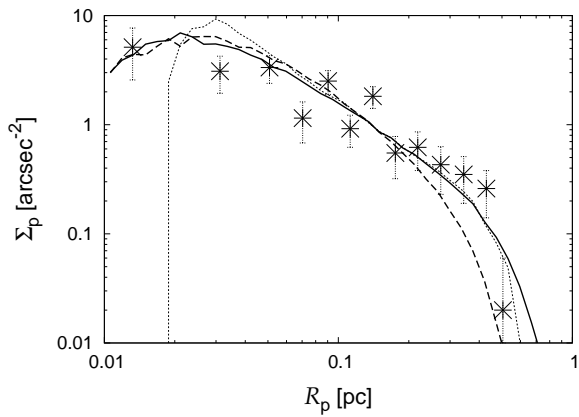}
\end{center}
\caption{Projected radial surface density profile,
$\Sigma_\mathrm{p}(R_\mathrm{p})$, of the evolved stellar disc
with initial value of index $\beta=-1.5$ within the interval
$a\in \langle 0.03\,\pc,\; 0.6\,\pc\rangle$ (solid line) and $\beta=-2$ for
$a\in \langle 0.04\,\pc,\; 0.4\,\pc \rangle\,$(dashed line). Thin dotted line
shows the initial state of the model with $\beta=-1.5$. The data points represent
projected density as determined by \cite{do2013} with scaling $1\mas = 0.04\,\pc$.
Note that the rightmost data point was not displayed in Figure~13 of the original
paper as the authors restricted themselves to the region below $12\mas$.
The modelled density profiles are normalised arbitrarily to match the
observational data.}
\label{fig:sigma}
\end{figure}

Distribution of the semi-major axes of the young stars in the GC cannot be
determined from the observational data without some \`a priori assumptions
about their configuration. Hence, instead of attempting to compare the output
of our model with (biased) deprojected structure of the system of young stars
in the GC, we plot projected surface density of the system in our model in
Figure~\ref{fig:sigma}. For this purpose, we choose angle $127\degr$ between
the line-of-sight and the symmetry axis of the disc which corresponds to
the inclination of the young stellar disc in the GC with respect to the
plane of the sky given by \cite{paumard2006}; other authors report only slightly
different values, e.g. $123\degr$ is the best fit of \cite{lb2003} who
identified the young stellar disc for the first time. Together with the
lines determined from our numerical models, we show projected surface density
of {\em all\/} the young stars in the GC (points in Figure~\ref{fig:sigma})
as given in \cite{do2013}. We do not consider reasonable to fit the models
against the surface density determined from
observations which suffers from
too large errors. Still, according to a by-eye comparison, the model with initial
density profile $\Sigma(R) \propto R^{-2}$ in $\langle 0.04\,\pc,\,
0.4\,\pc\rangle$ somewhat differs from the observational data. Hence, we
also show results for another model with $\beta=-1.5$ and $R\in \langle 0.03\,\pc,\,
0.6\,\pc\rangle$ which matches the observational data reasonably well.
Comparison of the two models further indicates that the profile below
$\approx 0.1\,\pc$ is not much affected by manipulation with the model
parameters, while these may be eventually
used to fit the observational data in the outer parts of the disc. In
this context, the most recently reported smaller outer radius of the disc
at $\approx0.13$ pc \citep{yel14} can also be straightforwardly accommodated.
\section{Discussion}
In previous sections, we have learned that the two-body relaxation can lead to
rapid changes of the density profile of a stellar disc around a SMBH.
The question now is whether this process cannot be marginalised
by some other, stronger effect. In Section~\ref{sec:theory} we have
indirectly argued that the two-body relaxation within the whole central star cluster
is likely to act on an orders of magnitude longer time-scale than the two-body
relaxation within the disc. Another forms of relaxation, in particular the vector
resonant relaxation, however, may be stronger opponents. The characteristic
time-scale of the vector resonant relaxation is \citep{rt1996,ha2006}
\begin{equation}
 t_\mathrm{RR,v}(a) \approx 0.6\left( \frac{\mbh}{\mstar} \right) \frac{P(a)}{N^{1/2}(a)}\;,
\end{equation}
where $P(a) = 2\pi\sqrt{a^3/G\mbh}$ is the orbital period of a star on orbit with
semi-major axis $a$ and $N(a)$ is the number of stars with semi-major axis $<a$. For
a star cluster with radial density profile $\rho(R) \propto R^\gamma$ this roughly
implies
\begin{eqnarray}
 t_\mathrm{RR,v}(R) &\bs\approx\bs& 2
 \left( \frac{\mbh / M_\mathrm{c,0.04}}{100} \right)^{1/2}
 \left( \frac{\mbh / \mstar}{10^6} \right)^{1/2} \times \nonumber \\
 & & 
 \left( \frac{\mbh}{4\!\times\!10^6\msun} \right)^{\!-1/2}
 \left( \frac{R}{0.04\,\pc} \right)^{-\gamma/2}\myr
\end{eqnarray}
with $M_\mathrm{c,0.04}$ denoting the mass of the nuclear star cluster below
$0.04\,\pc$. The process of vector resonant relaxation generally leads to changes
of orientations of the orbits, i.e. it tends to disrupt (thicken) the disc.
Similarly to the two-body relaxation within the disc, it is efficient mainly at
small radii. A full $N$-body model including the old star cluster is needed to
determine, whether the resonant relaxation among the young and old stars could
dominate over the two-body relaxation within the disc and, consequently, suppress
the evolution of its radial density profile.
\begin{figure}
\begin{center}
\includegraphics[width=\columnwidth]{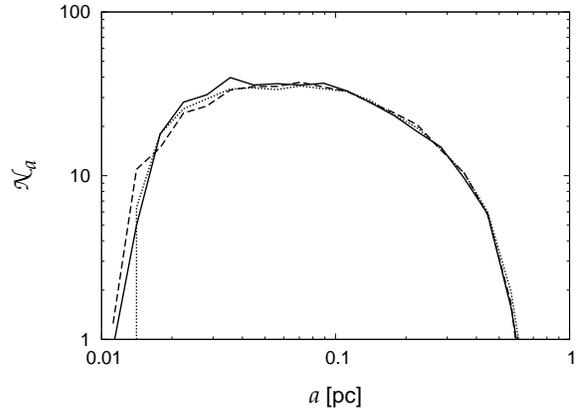}
\end{center}
\caption{Distribution of semi-major axes of stars forming the initially thin
stellar disc after several Myrs of dynamical evolution. Individual lines
correspond to model with standalone disc of stars in the dominating
potential of the central mass (dashed), disc in the potential of the central mass
superposed with smooth analytical potential mimicking the old star cluster
(dotted) and stellar disc embedded in a spherical cluster of 8000 gravitating
point-like particles (solid).}
\label{fig:na-nbody}
\end{figure}

A true full $N$-body model of the inner parsec of the GC is beyond
our current computational capabilities. Nevertheless, in order to obtain some
test of influence of the self-gravitating nuclear star cluster upon
the dynamical evolution of the disc, we introduce a simplified model. The
stellar disc is represented by 200 equal-mass stars of mass $100\,\msun$,
forming an initially thin disc with opening angle $4\degr$ which spans from
$0.04\,\pc$ to $0.4\,\pc$ with surface density $\Sigma(R) \propto R^{-2}$
(i.e. with the distribution of semi-major axes falling as $a^{-1}$). We followed
evolution of either the standalone disc in the dominating potential of the
central SMBH of mass $4\times10^6\msun$ or with the central Keplerian potential
superposed with a smooth potential of mass with radial density profile
$\rho(R) \propto R^{-7/4}$ (i.e. the \citealt{bw1976} solution) and, finally
the stellar disc embedded in a spherical cluster of 8000 stars of mass $50\,\msun$
with \cite{bw1976} distribution of semi-major axes and thermal ($\df_e
\propto e$) distribution of eccentricities. The evolved distributions of
semi-major axes are shown in Figure~\ref{fig:na-nbody}. The individual
models give results that are close enough to each other to make a conclusion
that the presence of the spherical old star cluster affects the evolution
of the stellar disc only marginally.

Beside the random torques from individual stars of the old star cluster, a systematic
influence of massive sources of gravity upon the young stellar disc could also
weaken the impact of two-body relaxation. Such sources have already been
discussed in the literature. Let us mention work of \cite{nayakshin2006} and
\cite{lb2009} who considered evolution of two stellar discs due to mutual
gravitational torques, works of \cite{ssk2009} and \cite{hsk2011,hsv2011}
taking into account torque exerted on the stellar disc by external massive gaseous
torus, or paper of \cite{mgh2013} dealing with the torques due to remnants
of the gaseous disc which gave birth to the young stars. A general statement
about the effect of systematic torques from massive, roughly axisymmetric
source of gravity upon the young stellar disc is that it leads to a change
of the orientation of the orbits of the stars in the disc. The strength of
the effect rapidly
decreases with the separation of the particular star and the perturbing source.
In accord with this, \cite{ssk2009,hsk2011,hsv2011} and \cite{mgh2013} concluded
that the perturbations under their consideration affect mainly the outer parts
of the disc, i.e. the two-body relaxation is likely to be dominant in the inner
parts of the stellar disc. In the case of two, mutually interacting discs of
similar radii, it is a matter of their mass, whether this interaction will lead to
a fast puff-up (i.e. lowering their densities) which could make the two-body
relaxation marginal.

\section{Conclusions}
\label{sec:conclusions}
We have shown that two-body relaxation may play an important role
in evolution of the radial density profile of a thin stellar disc
around a supermassive black hole given the disc is initially near-circular.
This assumption keeps the relative velocities of the stars in the disc
low enough to enable them to effectively interact.
In particular, considering the parameters
of the young nuclear cluster in the Galactic Centre, such a disc evolves on a
time-scale comparable to its age.

The question of evolution of semi-major axes of stars forming a self-gravitating
disc around a SMBH was also addressed by \cite{gual12} who did not find any significant
evolution of the radial density profile in their numerical models. This is likely
due to the fact that their distribution of initial eccentricities of the orbits
in the disc was peaked at value $\sqrt{\mean{e^2}}\approx0.3$. From the point of view
of dynamical evolution, the results of \cite{gual12} are in accord with our
findings as our 
calculations confirm that after $\sqrt{\mean{e^2}}\approx0.3$ is reached, the evolution
of the distribution of semi-major axes is negligible. In addition, \cite{lb2009}
also noticed accelerated energy relaxation in their models of stellar discs with small
initial eccentricities ($\mean{e} \approx 0.03$),
attributing it (as we do) to low relative velocities and high density.

Let us emphasise that it is important to be cautious when interpreting the
observational data with respect to the theoretical expectations about the
{\em initial\/} state of the system under consideration.
In particular, in this paper we have shown that starting from a rather simple initial
configuration, the dynamical evolution of the young stellar system in the GC over a few
millions of years may lead to a state compatible with recent observational
data. Within the framework of our model,
the characteristic time-scale of the disc evolution grows with the radial
distance from the centre (for reasonable initial profiles). We thus suggest that
only the outer parts of the disc ($\gtrsim0.1\,\pc$) may serve for
determining its initial state. In the inner parts, where the two-body
relaxation is more efficient, the distribution of semi-major axes tends to
be flat, implying the surface density decreasing roughly as $R^{-1}$ which
is remarkably close to the value reported by \cite{bse2009}.

We have further shown that large density at the inner edge of the disc leads
to its significant spreading towards
the centre. For $\beta=-2$ we obtained minimum value of semi-major axis
averaged over 20 realisations of the numerical model
$\mean{a_\mathrm{min}} \approx 0.019\,\pc$
at $t=6\,\myr$, i.e. less than one half of the initial inner radius of the
disc. For $\beta = -3$ we got even lower value, $\mean{a_\mathrm{min}}
\approx 0.01\,\pc$. Hence, we speculate that two-body relaxation may have
played an important role for the S-stars which are orbiting the supermassive
black hole with semi-major axes $0.005\,\pc \lesssim a \lesssim 0.04\,\pc$.
Although it seems that even a disc with the unrealistically steep initial
density profile does not provide sufficiently tight orbits, the two-body
relaxation may have worked on a longer period of time provided the S-stars originate
from some previous star formation episode and, at the same time, if not
placing the S-stars to their orbits, two-body relaxation may have helped to
do so other processes,
e.g. the tidal disruption of binaries on highly eccentric orbits around the
SMBH \citep[see e.g.][]{lbk2009,pg2010}.
\section*{Acknowledgments}
We appreciate kind help of the author of the \model{NBODY6} code, Sverre
Aarseth, with tuning the code for a better performance with the
additional Keplerian potential and Tuan Do for providing us with the data
plotted in Figure~\ref{fig:sigma}. This work was supported in part by the
National Science Foundation under Grant No. PHYS-1066293 and the hospitality
of the Aspen Center for Physics. We appreciate a financial support from
the Czech Ministry of Education through the grant LD12065 (the European
Action COST MP0905) and the Research Program MSMT0021620860.

\begin{appendix}
\section{Numerical tests}
\label{sec:appendix}
\begin{figure}
\begin{center}
\includegraphics[width=\textwidth]{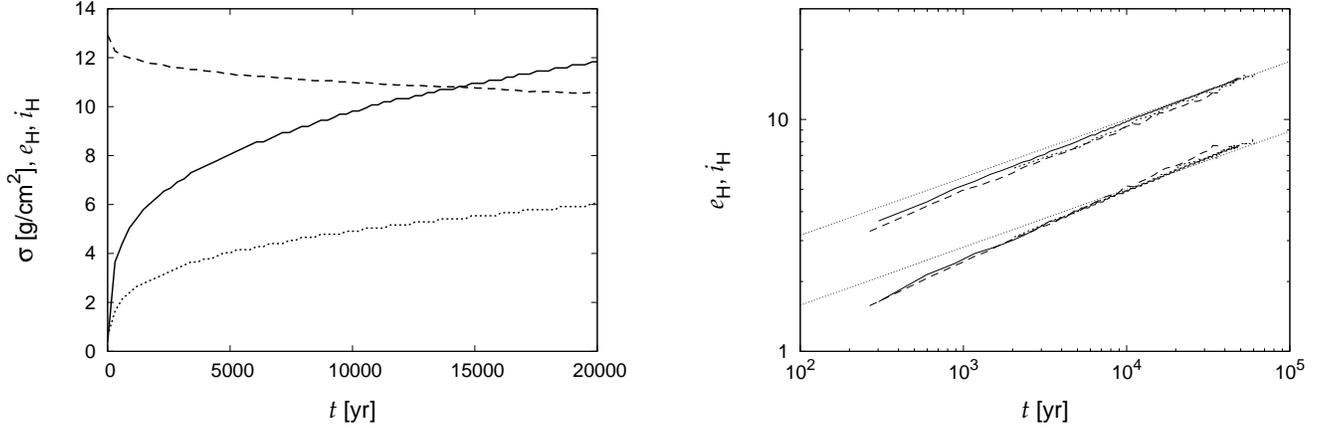}
\end{center}
\caption{Left: temporal evolution of surface density, $\sigma$ (dashed line), $\eH$
(solid line) and $\iH$ (dotted line) of a narrow
ring of 1024 bodies of mass $M = 5\times10^{-10}\mbh$ orbiting on nearly Keplerian
orbits around $\mbh$. The system is scaled to Solar system parameters for the sake
of comparison with Figure~2 of \cite{si2000}. Right: evolution of $\eH$ and $\iH$
for three different systems with time scaling according to formula~(\ref{eq:ts}).
Solid line stands for model with 1024 particles of mass $M = 5\times10^{-10}\mbh$,
dotted one for 256 particles of mass $M = 5.12\times 10^{-7}\mbh$ and dashed one
for the case of 40 particles of mass $M = 3\times10^{-6}$. Thin dotted lines show
$t^{1/4}$ dependence for comparison.}
\label{fig:si2000}
\end{figure}

In order to test robustness of the analytical results of \cite{si2000} and their
applicability to our configuration as well as to test reliability of the used
numerical code, we have performed several numerical tests.
First of them is slightly modified configuration \#1 introduced in \cite{si2000},
Section V.A. In particular, we integrated evolution of 1024 equal-mass particles
of mass $M = 5\times10^{-10}\mbh$ which formed initially a thin disc with constant
surface density ranging from $0.972\,R_0$ to $1.028\,R_0$. We set up initial
inclinations such that $\iH \equiv \sqrt{\mean{i^2}} (3\mbh/2M)^{1/3} = 1$. For
the sake of simplicity, we placed the stars on exactly circular orbits, which,
however, are in this case nearly indistinguishable from orbits with eccentricities
characterised by $\eH \equiv \sqrt{\mean{e^2}} (3\mbh/2M)^{1/3} = 2$ (i.e.
$\sqrt{\mean{e^2}} \approx 0.001$) which is the initial value of
\cite{si2000}. Left panel of Figure~\ref{fig:si2000} shows temporal evolution
of surface density, $\sigma$, and orbital elements in terms of $\iH$ and $\eH$ for
scaling $\mbh = \msun$ and $R_0 = 1\,\mathrm{AU}$. We see qualitative agreement
with Figure~2 of \cite{si2000}, including $\approx20\%$ decrease of the disc
surface density. In other words, there is no apparent discrepancy of our numerical
results with respect to both numerical and analytical results of \cite{si2000}.

Let us note that the numerical results can be directly rescaled
to arbitrary $\mbh$ and $R_0$ which implies time scaling $t \rightarrow t\,t_\mathrm{s}$
with $t_\mathrm{s} \equiv (R_0 / 1\,\mathrm{AU})^{3/2} \,/\; (\mbh / \msun)^{1/2}$.
Beside this trivial scaling, \cite{si2000} state that the temporal
evolution of $\iH$ and $\eH$ can be scaled for varying disc surface density and
mass of individual bodies (planetesimals in their case, stars in the context of
our paper) in the disc. In particular, they show that two discs with surface
densities $\sigma_1$ and $\sigma_2$ and disc bodies masses $M_1$ and $M_2$ will
give $i_\mathrm{H,1}(t) = i_\mathrm{H,2}(t_\mathrm{s}^\prime\,t)$ and, analogically,
$e_\mathrm{H,1}(t) = e_\mathrm{H,2}(t_\mathrm{s}^\prime\,t)$ with
\begin{equation}
 t_\mathrm{s}^\prime \equiv \frac{\sigma_1}{\sigma_2} \left(\frac{M_2}{M_1}
 \right)^{1/3}\;.
\label{eq:ts}
\end{equation}

We have tested this scaling by means of two additional models, one with 256 particles
of mass $M = 5.12\times 10^{-7}\mbh$ and the other one with 40 particles of mass
$M = 3\times10^{-6}\mbh$. In both cases, the disc of constant surface density spread
initially from $0.944\,R_0$ to $1.056\,R_0$. When rescaled to $\mbh = 4\times 10^6\msun$
and $R_0 = 0.04\,\pc$, both models have surface density comparable to the inner
parts of the disc described in Section~\ref{sec:model}. In addition, masses of
the bodies within the later model are approximately equal to the mean mass of stars
in the models presented in this paper. Right panel of Figure~\ref{fig:si2000} shows
temporal evolution of $\iH$ and $\eH$ for all of the three models introduced in this
Appendix with time scaled according to eq.~(\ref{eq:ts}). Note that, in evaluation
of $t_\mathrm{s}^\prime$, we have considered temporal evolution of $\sigma$ which we
determined numerically. For both $\iH$ and $\eH$, lines for the
three different models lie roughly on
the top of each other, indicating that (i) the theory of \cite{si2000} holds for
a wide range of the disc parameters and (ii) numerical results obtained using
of the {\tt NBODY6} code are in agreement with this theory.

Last, but not least, reliability of our numerical integrations is supported by the
energy conservation check -- in all models presented in this work, the total relative
energy error throughout the whole integration was $\lesssim 10^{-5}$.
\end{appendix}
\end{document}